\documentclass[twocolumn,prl,amsmath,amssymb,showpacs,
superscriptaddress,floatfix]{revtex4}

\usepackage{graphicx}
\usepackage {amsmath}
\DeclareMathOperator{\Tr}{Tr}

\begin{document}

\title{Cluster model of glass transition in simple liquids}

\author{N.M. Chtchelkatchev}
\affiliation{Institute for High Pressure Physics, Russian Academy
of Sciences, Troitsk 142190, Moscow Region, Russia}
\affiliation{L.D.\ Landau Institute for Theoretical Physics,
Russian Academy of Sciences, 117940 Moscow, Russia}

\author{V.N. Ryzhov}
\affiliation{Institute for High Pressure Physics, Russian Academy
of Sciences, Troitsk 142190, Moscow Region, Russia}

\author{T.I. Schelkacheva}
\affiliation{Institute for High Pressure Physics, Russian Academy
of Sciences, Troitsk 142190, Moscow Region, Russia}

\author{E.E. Tareyeva}
\affiliation{Institute for High Pressure Physics, Russian Academy
of Sciences, Troitsk 142190, Moscow Region, Russia}

\date{\today}

\begin{abstract}
On the basis of microscopic statistical mechanics of simple
liquids the orientational interaction between clusters consisting
of a particle and its nearest neighbors is estimated. It is shown that
there are ranges of density and temperature where the interaction
changes sign as a function of a radius of a cluster.
The model of
interacting cubic and icosahedral clusters is proposed and solved in
mean-field replica symmetric approximation. It is shown
that the glass order parameter grows
smoothly upon cooling,
the transition temperature being identified with the temperature
of the replica symmetry breaking. It is shown
that upon cooling a Lennard-Jones system, cubic clusters
freeze first. The transition temperature for icosahedral
clusters is about ten per cent lower. So the local
structure of Lennard-Jones glass in the vicinity of glass
transition should be most probably cubic.

\end{abstract}


\maketitle
Despite the growing interest to physical properties of liquids and glasses
(see, for example, for the recent reviews \cite{book,parisi,cugl}), nature of
structural glass transition
is still puzzling. While the experimental and phenomenological knowledge
of non-ergodic amorphous phases has considerably improved in last time \cite{book},
progress in first-principle statistical mechanical studies of structural glasses is much
more slow.

One of the promising ways to study the structural glasses is to explore the analogy
between the phase transitions in spin glasses (which are well understood now
\cite{cugl,parisi1}) and the   structural glass
transition. In some extent this is motivated by the fact that
the structure of the dynamical equations for the
appropriate correlation functions of super-cooled liquids and
$p$-spin spin glass model ($p\ge3$) are identical in the mean-field
(MF)
approximation \cite{cugl,KW,KT}, so this model may be used
to describe, at least qualitatively, the properties of structural
glasses.
The main obstacle in this way is the absence
in structural glasses of quenched disorder (in contrast to the spin glasses).
Spin glasses are microscopically quite different from liquids and
thus seem not suitable for their description. Furthermore, this
approach does not include any information on the local structure
of super-cooled liquids and glasses.
In this paper we propose the scenario where the analog of quenched disorder
in Lennard-Jones system appears in natural way allowing to apply
the methods of spin glass theory to investigation of the
structural glass transition.

The structure of supercooled liquids and glasses is incompletely
understood even for simplest systems. Frenkel \cite{frenkel} has suggested
a qualitative picture of local structure of a dense
supercooled liquid: he supposed that in small volumes it has a crystal-like
structure. On the other hand, as was argued by Frank \cite{frank},
an icosahedral cluster of 13 particles has a significantly lower
energy than more obvious arrangements having the symmetry of FCC
or
HCP crystals. It was inferred from the computer simulations of
a dense supercooled Lennard-Jones liquid \cite{st83} that
the local symmetry of simple liquids
is determined by defective or
fragmented icosahedral building blocks that exhibit a five-fold
symmetry. However, as was shown later, in a
supercooled Lennard-Jones liquid, small crystalline FCC clusters
nucleate following the temperature quench \cite{pat95}. It seems
that the local structure of a supercooled Lennard-Jones liquid is a
result of competition between FCC and icosahedral local
symmetries \cite{ryz}.

The concept of interacting clusters was used in phenomenological
theories of bond orientational order \cite{nt,hay1}
or "orientational melting" \cite{mp1,pr}. The microscopic
approach to description of the bond-orientational order (or
hexatic phase in two dimensions) was developed in
\cite{ryz,rt1,rt2,rt3}. In the present Letter we use this
approach to analyze the intercluster interaction and to introduce
the model of glass transition in a Lennard-Jones liquid which
elucidates the structural properties of the corresponding glass.

Our starting point is the expression for the free energy of the
system as a functional of a pair distribution function $g_2({\bf
r}_i,{\bf r}_0)$ which has the form \cite{ryz,rt1}:
\begin{eqnarray}
&&F/k_BT=\int d{\bf r}d{\bf r}_0 \rho g_2({\bf r},{\bf r}_0)\ln\left[\left(\lambda^3
\rho g_2({\bf r},{\bf r}_0)\right)-1\right]- \nonumber\\
&&-\sum_n \frac{\rho^{n+1}}{(n+1)!}\int S_{n+1}({\bf
r}_1...{\bf r}_{n+1})
g_2({\bf r}_1,{\bf r}_0)\cdots \nonumber\\
&&\cdots g_2({\bf r}_{n+1},{\bf
r}_0) d{\bf r}_1\cdots d{\bf r}_{n+1}d{\bf r}_0 -\nonumber\\
&&-\int \Phi({\bf
r}-{\bf r}_0)\rho g_2({\bf r},{\bf r}_0) d{\bf r}d{\bf r}_0. \label{1}
\end{eqnarray}
Here $S_{k+1}({\bf r}_1...{\bf r}_{k+1})$ is the irreducible cluster sum of
Mayer functions connecting (at least doubly) $k+1$ particles,
$\rho$ is the mean number density, $\Phi({\bf r}-{\bf r}_0)$ -
interparticle potential, $\lambda=h/(2\pi mk_BT)^{1/2}$.

In an isotropic liquid pair distribution function $g_2({\bf r},{\bf r}_0)$
depends on $|{\bf r}-{\bf r}_0|$ only: $g_2({\bf r},{\bf r}_0)=g(|{\bf r}-{\bf
r}_0|)$, where $g(r)$ is the radial distribution function. In the
state with bond orientational order rotational symmetry
of the pair distribution function is broken:
\begin{equation}
g_2({\bf r},{\bf r}_0)=g(|{\bf r}-{\bf r}_0|)+\delta g({\bf r},{\bf
r}_0). \label{2}
\end{equation}
$\delta g({\bf r},{\bf r}_0)$ has the symmetry of the local
environment of the particle at ${\bf r}_0$ and may be
approximately written in the form \cite{ryz,rt1}:
\begin{equation}
\delta g({\bf r},{\bf r}_0)=f(\Omega)\delta(r_s-|{\bf r}-{\bf
r}_0|). \label{3}
\end{equation}
Here $\Omega$ determines the direction of the vector ${\bf r}-{\bf
r}_0$, and $r_s$ is the size of the cluster, which is
approximately equal to the diameter of the first coordination
shell. Function $f(\Omega)$ gives the probability of cluster
orientation and may be expanded in a series in spherical
harmonics:
\begin{equation}
f(r,\Omega)=\sum_{l=0}^{\infty}\sum_{m=-l}^{l}f_{lm}(r)Y_{lm}(\Omega).
\label{4}
\end{equation}
In expansion (\ref{4}) only the terms corresponding to the local
cluster symmetry should be retained. Coefficients $f_{lm}$ are the
order parameters for the phase transition to the anisotropic
phase.

To estimate the intercluster interaction, let us expand the free
energy (\ref{1}) up to the second order in $\delta g({\bf r},{\bf
r}_0)$. Omitting the $\Omega$-independent terms, one has:
\begin{eqnarray}
\Delta F/k_bT&=&-\frac{1}{2}\int \Gamma({\bf r}_1,{\bf r}_0,{\bf
r}_2')\times\nonumber\\
&\times&\delta g({\bf r}_1,{\bf r}_0)\delta g({\bf r}_2,{\bf r}_0)
d{\bf r}_1 d{\bf r}_2, \label{5}
\end{eqnarray}
where \cite{ryz,rt1}:
\begin{eqnarray}
&&\Gamma({\bf r}_1,{\bf r}_1^0,{\bf r}_2)=\sum_{k\geq 1}\frac{\rho^k}
{(k-1)!}\int\,S_{k+1}({\bf r}_1...{\bf r}_{k+1})\times \nonumber\\
&&\times g(|{\bf r}_3-{\bf r}_1^0|)\cdots g(|{\bf r}_{k+1}-{\bf r}_1^0|)\,
d{\bf r}_3 \cdots d{\bf r}_{k+1}=\nonumber\\
&&=\sum_{l=0}^{\infty}\frac{4\pi}{2l+1}\Gamma_l(r,r')
\sum_{l=-m}^{l}Y_{lm}(\Omega_1)Y_{lm}^*(\Omega_2). \label{6}
\end{eqnarray}
The angles $\Omega_1$ and $\Omega_2$ determine the directions of the vectors
${\bf r}={\bf r}_1-{\bf r}_0$ and ${\bf r'}={\bf r}_2-{\bf r}_0$.

After substituting (\ref{3}) and (\ref{6}) in (\ref{5}) we obtain:
\begin{eqnarray}
&&\Delta F(r_s)/k_BT=-\frac{1}{2}\sum_{l=0}^{\infty}\frac{4\pi}{2l+1}
\Gamma_l(r_s,r_s)\times\nonumber\\
&&\times\int
\sum_{l=-m}^{l}Y_{lm}(\Omega_1)Y_{lm}^*(\Omega_2)
f(\Omega_1)f(\Omega_2)r_s^4d\Omega_1d\Omega_2=\nonumber\\
&&=-\frac{1}{2}\sum_{l=0}^{\infty} J_l(r_s)\sum_{l=-m}^l
|f_{lm}|^2. \label{7}
\end{eqnarray}
Here $J_l(r_s)=\frac{4\pi}{2l+1}r_s^4\Gamma_l(r_s,r_s)$.

Function $\Delta F(r_s)$ may be interpreted as the mean-field
orientational
interaction energy of the system of clusters having the size $r_s$. To get
the full energy of the system one should integrate (\ref{7}) over
the probability of finding the cluster with the size $r_s$ which
is given by the function $r^2g(r)$ in the
vicinity of the first maximum.

\begin{figure}[htb]
\includegraphics[width=80mm]{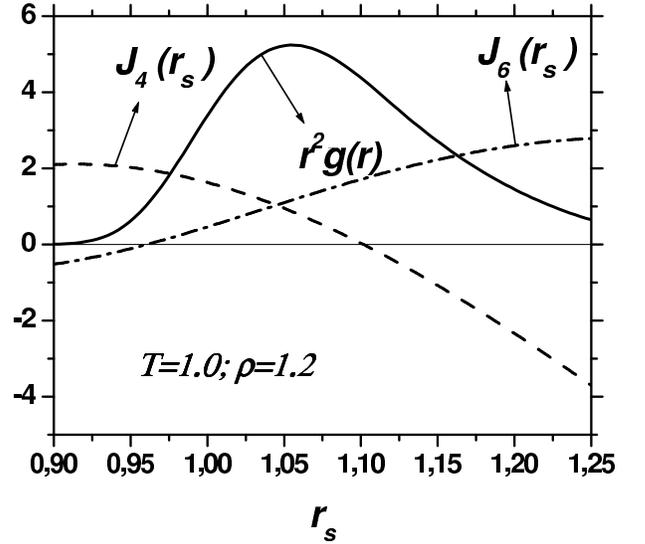}
\caption{\label{fig0} $J_l(r_s)$ for $l=4$
and $6$ along with $r_s^2g(r_s)$ in the vicinity of the first peak
as functions of $r_s$ for dimensionless temperature $T=1.0$ and density
$\rho=1.2$.}
\end{figure}

As was shown in \cite{ryz} there is
a simple approximation for $\Gamma({\bf r}_1,{\bf r}_1^0,{\bf r}_2)$
which gives rather good
results for the Lennard-Jones potential $\Phi_{LJ}(r)$:
\begin{equation}
\Gamma({\bf r}_1,{\bf r}_1^0,{\bf r}_2)=\rho
\left(\exp(-\Phi_{LJ}(|{\bf r}_1-{\bf r}_2|)/k_BT)-1\right). \label{71}
\end{equation}
Using Eq. (\ref{71}) one
can obtain the estimation for $J_l(r_s)$ as a function of
$r_s$. Fig.\ref{fig0} represents $J_l(r_s)$ for $l=4$
and $6$ along with $r^2g(r)$ in the vicinity of the first peak. It
is seen that $J_l(r_s)$ changes sign. From Eq. (\ref{7}) and Fig.
\ref{fig0} one can conclude that the frustration appears as a result of
variations in the sizes of clusters due to local density
fluctuations. Such kind of behaviour
leads to frustration which is analogous to that in spin glasses.

To study the transition in the system of interacting clusters
we introduce the simple lattice Hamiltonian::
\begin{equation}
H=-\frac{1}{2}\sum_{<i\neq
j>}\sum_{l=0}^{\infty}J^l_{ij}\sum_{m=-l}^l
U_{lm}(\Omega_i)U_{lm}^*(\Omega_j). \label{8}
\end{equation}
The functions $U_{lm}(\Omega_i)$ the lattice harmonics for the
space groups corresponding to the cluster symmetry. Taking into
account that $<U_{lm}(\Omega_i)>=f_{lm}$, one can see that in the
MF approximation the energy calculated from Eq. (\ref{8})
coincides with the intercluster energy (\ref{7}) under appropriate
choice of $J^l_{ij}$.

To simplify the problem we neglect in Hamiltonian (\ref{8}) all
the terms except for the ones corresponding to the unit
representation of the space group. Furthermore, we consider only
the cases $l=4$ and $l=6$ which represent the cases of cubic and
icosahedral symmetries.
This Ising-like model may be
called a "minimal" model. The Hamiltonian of the minimal model
has the form:
\begin{gather}
\label{one} H=-\frac{1}{2}\sum_{i\neq j}J_{ij}
\hat{U_i}\hat{U_j}.
\end{gather}
Functions $\hat U\equiv
U(\varphi,\theta)$ are the combinations of the spherical
harmonics. We will consider separately
symmetries of "simple" cube ($l=4, m=0,\pm 4$),
cube ($l=6, m=0,\pm 4$) and icosahedron ($l=6, m=0,\pm5$)
correspondingly \cite{hay1,Bredli}. The trace in this case is
defined as follows: $\Tr(\ldots)\equiv \int_0^{2\pi}d\varphi
\int_0^\pi d\cos(\theta)(\ldots)$. For example, for $l=4$ one has:
\begin{multline}
\hat U\equiv
U(\varphi,\theta)=\sqrt{\frac{7}{12}}\left\{Y_{40}(\varphi,\theta)+\right.
\\
\left.\sqrt{\frac{5}{14}}\left(Y_{44}(\varphi,\theta)+
Y_{44}(-\varphi,\theta)\right)\right\}
\end{multline}

The interactions $J_{ij}$ are such that the MF approximation
gives exact solution (an infinite-range interaction).
It is easily seen that in the minimal model (\ref{one}) without disorder in the
framework of the MF approximation there is a first order
phase transition to the state with bond orientational order.
From
Fig. \ref{fig0} it is clear that, as the first approximation, $J^l_{ij}$
may be random interaction with Gaussian probability
distribution.
\begin{gather}
\label{two} P(J_{ij})=\frac{1}{\sqrt{2\pi
J}}\exp\left[-\frac{(J_{ij}-J_0)^{2}}{ 2J^{2}}\right]
\end{gather}
with $ J=\tilde{J}/\sqrt{N}$ , $J_{0}=\tilde{J_0}/N$.

Then the free energy  in the replica-symmetric approximation is
 equal to \cite{Schelkacheva,4avtora}:
\begin{multline}
\label{four}
F=-NkT\biggl\{-\left(\frac{\tilde{J_0}}{kT}\right)\frac{m^2}{2}+
t^2\frac{q^2}{4}-t^2\frac{p^2}{4}+\\
\int_{-\infty}^{\infty}\frac{dz}{\sqrt{2\pi}}\exp\left(-\frac{z^2}{2}\right)\ln
\Tr\left[\exp\left(\hat\theta\right)\right]\biggr\}.
\end{multline}
Here $t=\widetilde{J}/k_BT$ and
$$\hat{\theta}=\left[zt\sqrt{q}+m\left(\frac{\tilde{J_0}}{kT}\right
)\right]\hat{U}+t^2\frac{p-q}{2}\hat{U}^2.$$

\begin{figure}[htb]
\includegraphics[width=80mm]{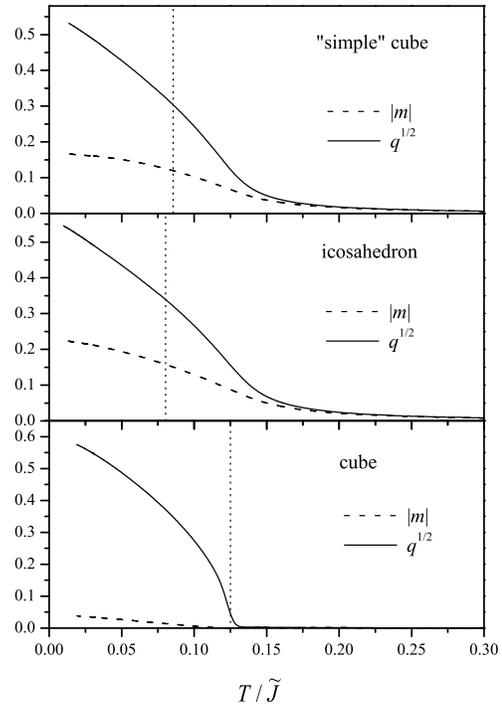}
\caption{\label{fig1} Order parameters $m$ and $q^{1/2}$ as
functions of $T/\tilde J$; $\tilde J_0=0$. Dotted vertical curve
shows the temperature $T_{_{A-T}}$ which separates stable and
unstable regions of the replica symmetric theory; the glass
transition occurs at $T_{_{A-T}}$.
"Simple" cube corresponds to $l=4, m=0,\pm 4$,
cube to $l=6, m=0,\pm 4$ and icosahedron to $l=6, m=0,\pm5$.}
\end{figure}

The order parameters are: $ m $ is the regular order parameter (an
analog of magnetic moment in spin glasses), $ q$ is the glass
order parameter and $p$ is an auxiliary order parameter. The
extremum conditions for the free energy ~\eqref{four} give the
following equations for these order parameters:
\begin{equation}
m=\overline{\langle \hat U\rangle},\,\,
p=\overline{\langle \hat U^2\rangle},\,\,
q=\overline{\langle \hat U\rangle^2} \label{prs},
\end{equation}
where $\langle\ldots\rangle=\Tr( \ldots e^{\hat\theta})/\Tr
e^{\hat\theta}$ and $\overline{(\ldots)}=\int_{-\infty}^{\infty}
\frac{dz}{\sqrt{2\pi}}e^{-z^2/2}[\ldots]$.
The temperature dependence of the order parameters is represented
in Fig.~\ref{fig1}. One can see that the glass phase grows
smoothly upon cooling. The similar behavior takes place in
a quadrupolar glass \cite{Schelkacheva,4avtora,h}. It is not clear
now whether the appearance of the nonzero bond-orientational order
is an artifact of the model or an inherent property of the glass
transition in the Lennard-Jones system.

\begin{figure}[htb]
\includegraphics[width=80mm]{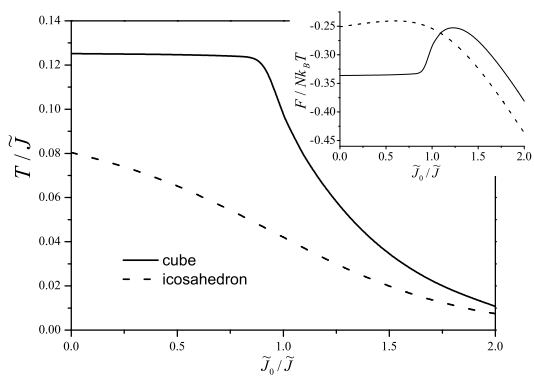}
\caption{\label{fig2} Evolution of the temperature $T_{_{A-T}}$
with $\tilde{J}_0/\tilde J$ is shown in the figure for the
symmetries of cube and icosahedron for $l=6$; below $T_{_{A-T}}$ the replica
symmetric MF solution becomes unstable. The free energy
$F(T_{_{A-T}}(\tilde J_0/\tilde J))/Nk_B T_{_{A-T}}$ is shown in
the inset.}
\end{figure}

We define the transition temperature as the replica symmetry
breaking temperature.
The replica symmetric solution is stable unless the replicon mode
energy $\lambda_{\rm repl}$ is nonzero \cite{A-T,4avtora}. For our
model we have
\begin{gather}
\lambda_{\rm repl}=1-t^2\overline{\langle\langle \hat U^2
\rangle\rangle^2},
\end{gather}
where $\langle\langle\ldots\rangle\rangle$ denotes the irreducible
correlator. The temperature $T_{_{A-T}}$ corresponding to
$\lambda_{\rm repl}=0$ defines the glass transition. In the
Fig.~\ref{fig2} we plot the dependence $T_{_{A-T}}$ on
$\widetilde{J}_0$ for $l=6$. From Fig.~\ref{fig2} one can conclude
that in the vicinity of the glass transition the cubic local
symmetry appears first and, taking into account the contribution
of the terms with $l=4$, is preferable for the Lennard-Jones
glass. This conclusion is in agreement with the result in Ref.
\cite{ryz}, where it was shown that the correlation length of the
cubic bond orientational order exceeds the icosahedral one.

From the condition of marginal stability \cite{cugl,KT} we also
evaluated numerically the so-called dynamical transition
temperature $T_{D}$ at $\tilde J_0=0$ and found that within the
accuracy of calculations $T_D$ and $T_{_{A-T}}$ coincide.
One can expect that in the minimal model (\ref{one}) full
replica symmetry breaking should occur (as in the
Sherrington-Kirkpatrick model \cite{parisi1}) but further
investigation is necessary to elucidate this question.

In summary, using the approach developed previously for study
of a bond-orientational order in simple liquids \cite{ryz}, we estimated
the orientational interaction between clusters consisting
of a particle and its nearest neighbors. It is shown that
there are ranges of density and temperature where the interaction
changes sign as a function of a radius of a cluster, the probability
of a radius being determined by the radial distribution function
of a liquid in the vicinity of the first peak. The model of
interacting cubic and icosahedral clusters was solved in
mean-field replica symmetric approximation. Due to absence
of reflection symmetry \cite{4avtora} the glass order parameter grows
smoothly upon cooling.
The transition temperature is identified with the temperature
replica symmetry breaking. It was shown
that upon cooling a Lennard-Jones system cubic clusters
freeze first. The transition temperature for icosahedral
clusters is about ten per cent lower. So the local
structure of Lennard-Jones glass in the vicinity of glass
transition should be most probably cubic.

We thank V. V. Brazhkin for stimulating discussions.
The work was supported by the Russian Foundation for Basic Research
(Grant No 02-02-16622 (VNR, NMC, and TIS),
Grant No 02-02-16621 (EET), and Grant No 03-02-16677 (NMC)).

\end{document}